\documentclass[prb,showpacs,aps,preprint]{revtex4}
\usepackage{pst-all,amsmath, amsthm, amsfonts,amssymb}
\newtheorem{theorem}{Theorem}[section]

\newtheorem{proposition}[theorem]{Proposition}
\newtheorem{corollary}[theorem]{Corollary}

\theoremstyle{definition}
\newtheorem{definition}[theorem]{Definition}
\newtheorem{example}[theorem]{Example}

\theoremstyle{remark}
\newtheorem{remark}[theorem]{Remark}

\numberwithin{equation}{section}

\newcommand{\cD}{\mathcal D}
\newcommand{\cF}{\mathcal F}
\newcommand{\cL}{\mathcal L}
\newcommand{\cS}{\mathcal S}

\newcommand{\C}{\mathbb{C}}
\newcommand{\LC}{\mathbb{L}}
\newcommand{\M}{\mathbb{M}}
\newcommand{\N}{\mathbb{N}}
\newcommand{\R}{\mathbb{R}}
\newcommand{\Mbil}[2]{\left<#1,#2\right>_\M}
\newcommand{\LCbil}[2]{\left<#1,#2\right>_\LC}

\newcommand{\bx}{\mathbf{x}}
\newcommand{\by}{\mathbf{y}}
\newcommand{\bp}{\mathbf{p}}

\newcommand{\LCphi}{\widetilde{\phi}}

\newcommand{\LComega}{\tilde{\omega}}

\newcommand{\LCcS}{\cS_{\partial_-}}
\newcommand{\LCp}{\tilde{p}}

\newcommand{\LCx}{\tilde{x}}
\newcommand{\LCy}{\tilde{y}}
\newcommand{\LCbp}{{\tilde{\bp}}}
\newcommand{\LCbx}{{\tilde{\bx}}}
\newcommand{\LCby}{{\tilde{\by}}}
\newcommand{\LCD}{{\widetilde{D}}}

\newcommand{\vac}{\!\left.|0\right>}
\newcommand{\bravac}{\left<0|\!\right.}

\newcommand{\pLCFT}{\cF_\LC^{\LCbx\to\LCbp}}
\newcommand{\LChat}[1]{{#1}^\sqcap}
\newcommand{\supp}{\mbox{supp}}

\newcommand{\tres}{|^*}
\newcommand{\WF}{\mbox{WF}}

\begin{document}
\title{On the problem of mass-dependence of the two-point function of the real scalar free
massive field on the light cone}
\author{Peter Ullrich}
\affiliation{Institut f\"ur Informatik, TU M\"unchen,
Boltzmannstra\ss e 3, D-85748 Garching, Germany}
\altaffiliation[Also at ]{Institut f\"ur Physik, Universit\"at Regensburg.}
\email{ullrichp@in.tum.de}

\author{Ernst Werner}
\affiliation{Institut f\"ur Physik, Universit\"at Regensburg,
Universit\"atsstra\ss e 31, 
D-93040 Regensburg, Germany}
\email{ernst.werner@physik.uni-regensburg.de}
%\date{\today}

\begin{abstract}
We investigate the generally assumed inconsistency in light cone quantum field theory that
the restriction of a massive, real, scalar, free field to the nullplane
$\Sigma=\{x^0+x^3=0\}$ is independent of mass \cite{LKS}, but the restriction of the 
two-point function is mass-dependent (see, e.g., \cite{NakYam77, Yam97}). 
We resolve this inconsistency by
showing that the two-point function has no canonical restriction to $\Sigma$
in the sense of distribution theory.
Only the so-called tame restriction of the two-point function,
which we have introduced in \cite{Ull04sub}, exists. Furthermore, we show that this
tame restriction is indeed  independent of mass. Hence the inconsistency is induced
by the erroneous assumption that the two-point function has a (canonical) restriction 
to $\Sigma$.
\end{abstract}
\maketitle
%
%-----------------------------------------------------------------------------
\section{Introduction}
%-----------------------------------------------------------------------------
%
\noindent
Let $\phi(x)$ be the real, scalar, free quantum field of mass $m>0$, and let $\vac$ denote
the (unique) vacuum state. The (Wightman) $n$-point functions (or vacuum expectation
values) are defined by $W_n(x_1,\ldots,x_n)=\bravac\phi(x_1)\cdots\phi(x_n)\vac$
($n\in\N$). 
Since $\phi$ is a free field, the two-point function $W_2(x,y)$ is explicitly given by
$W_2(x,y)=-i D^{(-)}_m(x-y)$, where $D^{(-)}_m(x)$ is the negative frequency 
Pauli-Jordan function
$$
	D^{(-)}_m(x)=\frac{-1}{i(2\pi)^3}\int\frac{d^3\bp}{2\omega(\bp)}\,
			e^{-i(\omega(\bp)x^0+\bx\cdot\bp)}.
$$
Treating the field $\phi$ in the framework of light cone quantization, the canonical
commutator relation reads
\begin{equation}\label{eq:LC-commutator}
	[\LCphi(\LCx)\LCphi(\LCy)]_{x^+=y^+=0}=
		\frac{1}{4i}\epsilon(x^- - y^-)\delta(\bx_\bot - \by_\bot),
\end{equation}
where we have introduced {\em light cone coordinates}
$$
	\LCx=(x^+,\LCbx)=(x^+,\bx_\bot,x^-)=\kappa(x^0,x^1,x^2,x^3)
$$
by 
$$
	x^+=(1/\sqrt 2)(x^0+x^3), \quad \bx_\bot=(x^1,x^2), \quad 
	x^-=(1/\sqrt 2)(x^0 - x^3).
$$
Furthermore,  $\LCphi(\LCx)=\phi(\kappa^{-1}(\LCx))$ denotes the transformed field. There
is a generally alleged inconsistency in light cone quantum field theory 
(see for example \cite{NakYam77, Yam97}) which we explain now in detail: Using
the commutator relation \eqref{eq:LC-commutator} one formally obtains the equation
\begin{equation}\label{eq:intro2}
	\bravac \LCphi(\LCx)\LCphi(0)\vac_{x^+=0}=
	\frac{1}{2\pi}\int_{p^+>0}\frac{dp^+}{2p^+}e^{-p^+x^-}\delta(\bx_\bot),
\end{equation}
where the right-hand side obviously does not depend on the mass. Since
$W_2(x,y)=-i D^{(-)}_m(x-y)$, \eqref{eq:intro2} should be equal to $-i$ times the restriction
of $\LCD^{(-)}_m(\LCx)$ to $x^+=0$, where 
$\LCD^{(-)}_m(\LCx)=D^{(-)}_m(\kappa^{-1}(\LCx))$ denotes the 
negative frequency Pauli-Jordan function transformed to light cone coordinates. 
In (3+1)-dimensional Minkowski space $D^{(-)}_m(x)$ has the following 
explicit representation \cite{Bog1}
\begin{equation}\label{eq:LCDneg_explicit}
	D^{(-)}_m(x)=\lim_{\substack{\xi\to 0\\ \xi\in V^+}}
		\frac{i m^2}{4\pi^2}h(-m^2(x-i\xi)^2),
\end{equation}
where $h(\zeta)=K_1(\sqrt\zeta)/\sqrt\zeta$,
$K_1$ is the modified Bessel function of second kind and the branch of 
$\sqrt\zeta$ is taken to be positive for $\zeta>0$.
One seemingly obtains a contradiction by transforming formally the right-hand
side of \eqref{eq:LCDneg_explicit} to LC-coordinates and putting $x^+=0$, because then
the right-hand side remains dependent on the mass $m$. However, as we will see later,
the formal manipulations at the right hand side of \eqref{eq:LCDneg_explicit}
are ill-defined, since $D^{(-)}_m(x)$ has no (canonical) restriction
to $\{x^0+x^3=0\}$. More precisely, the operations of taking the limit 
$\xi\to 0$ ($\xi\in V^+$) (in $\cS'(\R^4)$ -- the space of tempered distributions)
and putting $x^+=0$ do not commute in \eqref{eq:LCDneg_explicit}.
%
%-----------------------------------------------------------------------------
\section{Notations and Conventions}
%-----------------------------------------------------------------------------
%
Already in the introduction we have introduced light cone coordinates $\LCx=\kappa(x)$
by using the Kogut-Soper convention \cite{BrosPins}, where $x=(x^\mu)$ are 
Minkowski coordinates. As usual in light cone physics one writes
$$
	\LCx=(x^+,\LCbx)=(x^+,\bx_\bot,x^-), \qquad \bx_\bot=(x^1,x^2).
$$ 
The Minkowski bilinear form 
$\Mbil{x}{y}=x^\mu x_\mu=x^\mu g_{\mu\nu}x^\nu$, where 
$(g_{\mu\nu})=\mbox{diag}(1,-1,-1,-1)$ is the usual Minkowski metric, transforms to
the so-called LC-bilinear from 
$$
	\LCbil{\LCx}{\LCy}=
	\Mbil{\kappa^{-1}(\LCx)}{\kappa^{-1}(\LCy)}=
	x^+y^- + x^-y^+ - \bx_\bot\cdot\by_\bot 
	\qquad (\bx_\bot\cdot \by_\bot=x^1y^1+x^2y^2)
$$
when going over from Minkowski- to light cone coordinates; hereby 
$\kappa:\R^4\to\R^4$ is the linear transformation from Minkowski- to light cone coordinates

We also use light cone coordinates $\LCp=\kappa(p)$ in momentum space. However,
since $x^+$ is the time variable in light cone physics, and in
$\LCbil{x}{p}$ $x^+$ is multiplied by $p^-$, the variable $p^-$ takes on the role of
energy and $\LCbp=(p^+,\bp_\bot)$ is the (light cone) spatial momentum. 
Hence the light cone variable $\LCp$ is split into $\LCp=(\LCbp,p^-)$
with $\LCbp=(p^+,\bp_\bot)$ in contrast to the LC-space-time variable $\LCx$ which
we have split into $\LCx=(x^+,\LCbx)$ with $\LCbx=(\bx_\bot,x^-)$. Here a little bit
care is needed. 

Throughout this paper we denote by $\Sigma_\tau$ $(\tau\in\R)$ the linear subspace
$$
	\Sigma_\tau=\{x\in\R^4:(1/\sqrt 2)(x^0+x^3)=\tau\},
$$
where, especially for $\tau=0$, we set $\Sigma=\Sigma_0$.
Note that in light cone coordinates $\Sigma_\tau$ reads $\{x^+=\tau\}$, i.e.,
$\kappa(\Sigma_\tau)=\{\LCx\in\R^4:x^+=\tau\}$.

If $U\subset\R^m$ is an open set, we denote by $\cD(U)$ the (complex) vector space
consisting of all (complex-valued) smooth, i.e., $C^\infty$ functions on $U$ with compact
support. On $\cD(U)$ one defines a topology which makes $\cD(U)$ into a complete
locally convex space \cite{Hoer1, Rudin_func}, the dual space $\cD'(U)$ is called the
space of {\em distributions}. One canonically identifies $\cD(U)$ with a
subspace of $\cD'(U)$, i.e., we may assume $\cD(U)\subset\cD'(U)$, and, with respect
to the weak$^*$-topology \cite{Rudin_func} on $\cD'(U)$, $\cD(U)$ is even dense in 
$\cD'(U)$. Along with $\cD(\R^m)$ one introduces the Schwartz space $\cS(\R^m)$ 
of rapidly decreasing functions and defines on $\cS(\R^m)$ a topology which makes 
$\cS(\R^m)$ into a Fr\'echet space. The dual space $\cS'(\R^n)$ is called the space of
{\em tempered distributions (or generalized functions)}
\cite{Hoer1, Rudin_func, Bog1}. As in the case of distributions we may assume
$\cS(\R^m)\subset\cS'(\R^m)$, and $\cS(\R^m)$ is dense in $\cS'(\R^m)$ where
$\cS'(\R^m)$ is endowed with the weak$^*$-topology. Notice, that 
$\cD(\R^m)\subset\cS(\R^m)$, but the topology of $\cD(\R^m)$ is finer than the subspace
topology induced by $\cS(R^m)$. One usually identifies the subspace of distributions
($\in\cD'(\R^m)$) which admit a linear, continuous extension to $\cS(\R^m)$ with
$\cS'(\R^m)$.
%
%------------------------------------------------------------------------------------------
\section{Canonical restriction and wave front set}
%------------------------------------------------------------------------------------------
%
In this section we summarize some well-known results 
from distribution theory \cite{Hoer1} which will be needed in the sequel. Assume
$U\subset \R^m$ and $V\subset\R^n$ are open sets and $a\in U$ is fixed. Then the
restriction of a (classical) function $\phi(x,y)$ on $U\times V$ to $\{x=a\}$ can 
be viewed as the result of a 
pullback operation. More precisely, the restriction $y\mapsto\phi(a,y)$ equals
the pullback $\iota_a^*\phi=\phi\circ\iota_a$, where $\iota_a:V\to U\times V$,
$y\mapsto(a,y)$. Hence, the restriction operation is a special case of the pullback
operation which is generally defined by $\phi\mapsto f^*\phi=\phi\circ f$ where
$f:X\to Y$ is some (fixed) map and $\phi$ is a function on $Y$;  $f^*\phi$ is
called the {\em pullback} of $\phi$ with respect to $f$. Especially if  $f$ is
smooth, i.e., $C^\infty$, and $X$, $Y$ are open sets then $f^*$ maps $\cD(Y)$ into
$\cD(X)$; moreover, $f^*$ is linear and continuous. From distribution theory it follows
\cite{Hoer1} that it is impossible to extend $f^*$ (sequentially) continuously to
a linear map from $\cD'(Y)$ into $\cD'(X)$ unless conditions are imposed on $f$.
Only if $f$ is a submersion, i.e., the differential $d_xf$ is surjective for every $x\in X$,
a sequentially continuous extension of $f^*$ exists \cite{Hoer1}. However,
in the situation of the restriction operation the map $\iota_a:V\to U\times V$ is by no
means a submersion -- this is easily seen by comparing the dimensions of the associated
tangent spaces. Hence the extension of the restriction operation from classical functions to
distributions is more subtle. The most crucial ingredient in this case is the so-called 
wave front set which takes control on the singularities of a distribution. For details on
the wave front set we refer the reader to \cite{Hoer1}. The following theorem 
from \cite{Hoer1} determines the right subspace of $\cD'(Y)$ to which 
the pullback operation $f^*$ can be extended from $C^\infty(Y)$ when $f:X\to Y$ is 
generally a $C^\infty$ map.  Thereby Hoermander introduces for any conic subset
$\Gamma$ of $Y\times(\R^n\setminus 0)$ the subspace
$$
	\cD'_\Gamma=\{\phi\in\cD'(Y):\supp(\phi)\subset\Gamma\}
$$
which, however, carries a stronger topology than the subspace topology induced by
$\cD'(Y)$. Furthermore, one also needs to define the subspace
$$
	N_f=\{(f(x),\eta)\in Y\times\R^n : (d_x f)^t\eta=0\}
$$
which is called the {\em set of normals} of $f$.
\begin{theorem}[\cite{Hoer1}, Thm.\ 8.2.4 ]\label{thm:pullback}
Let $X\subset\R^m$ and $Y\subset\R^n$ be open subsets and let $f:X\to Y$ be
a $C^\infty$ map. Then the pullback $f^*\phi$ can be defined in one and only one way
for all $\phi\in\cD'(Y)$ with
$$
	N_f\cap\WF(\phi)=\emptyset
$$
so that $f^*\phi=\phi\circ f$ when $\phi\in C^\infty(Y)$. Moreover, for any closed conic
set $\Gamma$ of $Y\times(\R^n\setminus 0)$ with $\Gamma\cap N_f=\emptyset$ we
have a continuous map
$$
	f^*:\cD'_\Gamma(Y) \to \cD'_{f^*\Gamma}(X),
$$
where
$$
	f^*\Gamma=\{(x,(d_xf)^t\eta):(f(x),\eta)\in\Gamma\}.
$$
\end{theorem}
From Theorem~\ref{thm:pullback} one immediately obtains
\begin{equation}\label{eq:pullback_WF}
	\WF(f^*\phi)\subset  f^*\WF(\phi)
\end{equation}
whenever $N_f\cap\WF(\phi)=\emptyset$. Since the pullback operation is a (contravariant)
functor, i.e., $(g\circ f)^*=f^*\circ g^*$ ($g:V\to W$), one obtains from 
\eqref{eq:pullback_WF}:
\begin{corollary}\label{cor:pullback_diffeo}
Let $f:X\to Y$ ($X,Y\subset\R^m$) be a $C^\infty$ diffeomorphism. Then
$$
	\WF(f^*\phi)=f^*\WF(\phi)
$$
for all $\phi\in\cD'(Y)$. (Notice that $N_\lambda=Y\times\{0\}$.)
\end{corollary}
The definition of the canonical restriction of a distribution rests on the above theorem.
One just applies the theorem to the case when $f$ is the map 
$\iota_a:V\to U\times V$. Note that, by this, not all distributions of $\cD'(U\times V)$ have a 
canonical restriction to $\{x=a\}$.
\begin{definition}
Let $U\subset \R^m$, $V\subset\R^n$ be open subsets and $a\in U$. Then we say that
$\phi(x,y)\in\cD'(U\times V)$ has a (canonical) restriction to
$\{x=a\}$ if
$$
	N_{\iota_a}\cap\WF(\phi)=\emptyset,
$$
where $\iota_a:V\to U\times V$, $y\mapsto(a,y)$, and call 
$\phi|_{x=a}(y)=\phi(a,y)=\iota_a^*\phi(y)\in\cD'(V)$ 
the {\em canonical restriction} of $\phi(x,y)$ to $\{x=a\}$.
\end{definition}
\begin{remark}
(a) One easily computes the set of normals of $\iota_a$:
\begin{equation}\label{eq:restriction_condition}
	N_{\iota_a}=(\{a\}\times\R^n)\times(\R^m\times\{0\})
\end{equation}
Hence $\phi(x,y)$ has a canonical restriction to $\{x=a\}$ if and only if
$$
	((\{a\}\times\R^n)\times(\R^m\times\{0\}))\cap\WF(\phi)=\emptyset.
$$

(b) If $\phi(x,y)\in\cD'(U\times V)$ has a restriction to $\{x=a\}$ then
also any $\partial^\alpha_x \partial^\beta_y\phi(x,y)$ ($\alpha$, $\beta$ multi-indices)
has a restriction to $\{x=a\}$ -- by \cite{Hoer1}, 
$\WF(\partial^\alpha_x \partial^\beta_y\phi)\subset\WF(\phi)$.

(d) Since the wave front set is a closed set one easily finds that  
if $\phi(x,y)\in\cD'(U\times V)$ has a restriction to $\{x=a\}$ then there is an open 
neighborhood $U'\subset U$ of $a$ such that $\phi(x,y)$ has a restriction to $\{x=a'\}$ for
all $a'\in U'$. 
\end{remark}
% ------------------------------- Figure -------------------------------------------------------------------------
\begin{figure}[t]
\vspace{3cm}
\input{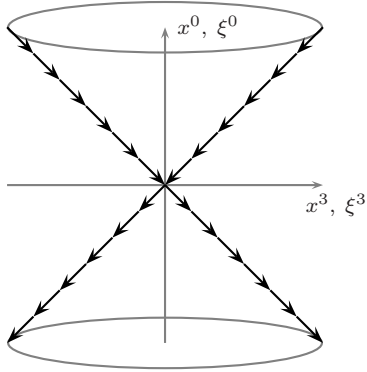}
\vspace{2.3cm}
\caption{The wave front set of $D^{(-)}_m$}
\label{fig:WF_neg_Pauli}
\end{figure}
% -----------------------------------------------------------------------------------------------------------------
\begin{remark}
Condition \eqref{eq:restriction_condition} in the definition of the restriction of a distribution
looks a little bit artifical, however, one can show (see, e.g., \cite{PhD}) that 
$\phi(x,y)\in\cD'(U\times V)$ has a restriction to $\{x=a\}$ if and only if, sufficiently
close to $x=a$,
$\phi(x,y)$ is $C^\infty$-dependent on $x$ as a parameter, i.e., there is an
open neighborhood $U'\subset U$ of $a$ and a family $\phi_x(y)\in\cD'(Y)$
$(x\in U')$ such that $U'\to\C$, $x\mapsto (\phi_x,g)$ is $C^\infty$
for every $g\in\cD(Y)$ and
$$
	(\phi(x,y),f(x)g(y))=\int_{U'}(\phi_x,g) f(x) dx
$$
for all $f(x)\in\cD(U')$, $g(y)\in\cD(V)$; if this is the case $\phi(a,y)=\phi_a(y)$.
\end{remark}
\begin{example} 
The Pauli-Jordan function 
$$
	D_m(x)=\frac{1}{i(2\pi)^3}\int d^4p\, \epsilon(p^0)\delta(p^2-m^2)\,
		e^{i\Mbil{p}{x}}\in\cS'(\R^4)
$$
has a canonical restriction to $\{x^0=0\}$. Moreover, $D_m(x)$ is even a fundamental 
solution of the Klein-Gordon operator, i.e.,
$$
	D_m(0,\bx)=0,\qquad (\partial_{x^0}D_m)(0,\bx)=\delta(\bx).
$$
That $D_m(x)$ has a restriction to $\{x^0=0\}$ can either be seen by considering the
wave front set of $D_m$ or, more explicitly, by showing that $D_m(x^0,\bx)$ is
$C^\infty$-dependent on $x^0$ as a parameter,
\footnote{This is indeed true for every solution of the Klein-Gordon equation since
the Klein-Gordon operator is hypoelliptic with respect to $x^0$, 
see \cite{Garding1958, Garding1961, Ehrenpreis1962}.} 
where
$$
	(D_m)_{x^0}(\bx)=\frac{1}{(2\pi)^3}\int \frac{d^3\bp}{\omega(\bp)}
		\sin(\omega(\bp)x^0)e^{-i\bp\cdot\bx} 
$$
Also the positive- and negative-frequency parts $D^{(\pm)}_m(x)$ have restrictions to
$\{x^0=\tau\}$ ($\tau\in\R$). In \cite{ReedII} the wave front set of $D^{(-)}_m(x)$ is 
explicitly determined:
\begin{align*}
    \WF(D^{(-)}_m & (x))=W^{(-)}_0\cup W^{(-)}_+\cup W^{(-)}_-,
    \text{with}\\
    & W^{(-)}_0=\{(0,\xi):\xi\in\Gamma^-\setminus 0\},~
      W^{(-)}_\pm = \{(\xi,\mp\lambda\xi):\xi\in\Gamma^\pm\setminus 0,~
      \lambda>0)\}.
\end{align*}
Thus $N_{\iota_\tau}\cap\WF(D^{(-)}_m)=\emptyset$ for all $\tau\in\R$. Since
$D^{(+)}_m(x)=-D^{(-)}_m(-x)$, and hence $\WF(D^{(+)}_m)=-\WF(D^{(-)}_m)$,
the same holds true for $D^{(+)}_m$. 
In Figure~\ref{fig:WF_neg_Pauli}
we have illustrated the wave front set of $D^{(-)}_m$. Each element $(x,\xi)$ of 
$\WF(D^{(-)}_m)$ is represented by a pointed vector with base point $x$ and
unit vector in direction of $\xi$.
\end{example}
So far we have only defined the restriction of a distribution $\phi(x,y)$ to a hyperplane of 
the form $\{x=a\}$ ($a\in\R^m$). However, any smooth submanifold of $\R^m$ can
be described locally in such a manner by using appropriate charts. 
Let $\Sigma_\tau=\{(1/\sqrt 2)(x^0+x^3)=\tau\}$ ($\tau\in\R$)
and $\kappa$ the linear transformation to light cone coordinates, then
$\Sigma_\tau=\{\kappa^{-1}(\LCx):x^+=\tau\}$. Hence we define:
\begin{definition}
A distribution $\phi(x)\in\cD'(\R^4)$ has a (canonical) restriction to 
$\Sigma_\tau$ ($\tau\in\R$)
if $\kappa_*\phi(\LCx)=(\phi\circ\kappa^{-1})(\LCx)$ has a (canonical) restriction to
$\{x^+=\tau\}$. In this case we call 
$\phi|_{\Sigma_\tau}=\kappa_*\phi(0,\LCbx)\in\cD'(\R^3)$
the {\em (canonical) restriction} of $\phi$ to $\Sigma_\tau$.
\end{definition}
\begin{remark}
More generally, one can define the restriction of a distribution 
$\phi(x_1,\ldots,x_r)\in\cD'(\R^4\times\cdots\times\R^4)$ to 
$\Sigma_{\tau_1}\times\ldots\times\Sigma_{\tau_r}$
as the restriction of $\phi(\kappa^{-1}(\LCx_1),\ldots,\kappa^{-1}(\LCx_r))$
to $\{x_1^+=\tau_1,\ldots,x_r^+=\tau_r\}$, where $\LCx_i=(x_i^+,\LCbx_i)$ 
and $\tau_i\in\R$ ($i=1,\ldots,r$).
\end{remark}

\begin{remark}\label{rem:alt_sigma_res}
If we denote by $\lambda:\R^3\to\R^4$, 
$\LCbx=(x^1,x^2,x^-)\mapsto(x^-/\sqrt 2,x^1,x^2,-x^-/\sqrt 2)$ 
then $\lambda(\R^3)=\Sigma=\{x^0+x^3=0\}$
and $\lambda=\kappa^{-1}\circ\tilde\iota_0$, where $\tilde\iota_0(\LCbx)=(0,\LCbx)$.
Hence
$$
	\lambda^*\phi=(\kappa^{-1}\circ\tilde\iota_0)^*\phi=\tilde\iota^*(\kappa_*\phi)=
	\phi|_\Sigma,
$$
i.e., $\phi|_\Sigma$ is the pullback of $\phi$ with respect to $\lambda$. Notice that
$\lambda$ is a smooth parametrization of $\Sigma$, but $\Sigma$ has infinitely many. 
However, if $\mu$ is another smooth
parametrization of $\Sigma$ then $\lambda=\mu\circ(\mu^{-1}\circ\lambda)$, where
$\mu^{-1}\circ\lambda$ is a $C^\infty$ diffeomorphism from $\R^3$ by $\R^3$. Hence,
by Corollary~\ref{cor:pullback_diffeo},
$\lambda^*\phi=$ exists if and only $\mu^*\phi$ exists, and in this case
$\lambda^*\phi$ and $\mu^*\phi$ differ only by multiplication of a smooth function -- the
determinant of the Jacobi matrix of $\mu^{-1}\circ\lambda$. 
\end{remark}
%------------------------------------------------------------------------------------------
\section{Nonexistence of the restriction of the two-point function to the nullplane}
%------------------------------------------------------------------------------------------
%
%\begin{lemma}\label{lem:LCWF}
%Let $\LCGamma^\pm=\{\LCxi\in\LC^{1+n}:\LCxi^2=0,~\pm\LCxi^+>0\}$,
%then
%\begin{itemize}
%\item[\textrm{(i)}]
%    $\WF(\LCD^{(+)}_m)=\LCW^{(+)}_0\cup \LCW^{(+)}_+\cup \LCW^{(+)}_-$
%	where\\
%	$\LCW^{(+)}_0=\{(0,\LCxi):\LCxi\in\LCGamma^+\setminus 0\},~
%      \LCW^{(+)}_\pm =
%      \{(\LCxi,\mp\lambda\LCxi):\LCxi\in\LCGamma^\mp\setminus 0,~
%      \lambda>0)\}$\\
%and\\
%    $\WF(\LCD^{(-)}_m)=\LCW^{(-)}_0\cup \LCW^{(-)}_+\cup \LCW^{(-)}_-$
%    where\\
%    	$\LCW^{(-)}_0=\{(0,\LCxi):\LCxi\in\LCGamma^-\setminus 0\},~
%      \LCW^{(-)}_\pm =
%      \{(\LCxi,\mp\lambda\LCxi):\LCxi\in\LCGamma^\pm\setminus 0,~
%      \lambda>0)\}$.
%\item[\textrm{(ii)}] $\WF(\LCD_m)=\WF(D^{(+)}_m)\cup\WF(D^{(-)}_m)$
%\end{itemize}
%\end{lemma}
%We have illustrated the wave front set
%of $\LCD^{(-)}_m$ in Figure~\ref{fig:WF_neg_LCPauli} , 
%where $(\LCx,\LCxi)\in\WF(\LCD^{(-)}_m)$ is represented by
%a pointed vector with base point at position $\LCx$ and (unit-) vector
%in direction $\LCxi$.
%\begin{figure}
%\vspace{3cm}
%\input{fig5.tex}
%\vspace{2.3cm}
%\caption{The wave front set of $\LCD^{(-)}_m$}
%\label{fig:WF_neg_LCPauli}
%\end{figure}
Since we have explicit knowledge of the wave front set of $D^{(-)}_m$ it it easy now
to show that the two-point function $W_2(x,y)$ has no (canonical) restriction to
$\Sigma\times\Sigma$.
% ------------------------------- Figure 2 ----------------------------------------------------------------------
\begin{figure}[t]
\vspace{3cm}
\input{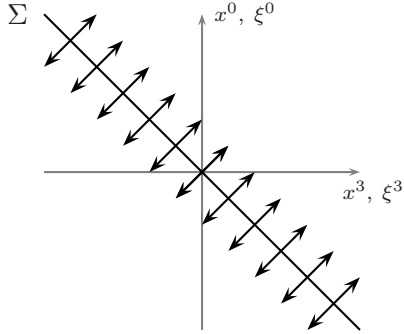}
\vspace{2.3cm}
\caption{The set of normals $N_\lambda$}
\label{fig:NormalSetLambda}
\end{figure}
% -----------------------------------------------------------------------------------------------------------------
\begin{theorem}\label{thm:nonexistence_res_two-point}
Let $W_2(x,y)\in\cD'(\R^4\times\R^4)$ denote the two-point function of the real scalar
free massive field. Then $W_2(x,y)$ has no canonical restriction to 
$\Sigma\times\Sigma=\{x^0+x^3=y^0+y^3=0\}$.
\end{theorem}
\begin{proof}
Since $W_2(x,y)=-iD^{(-)}_m(x-y)$ it is enough to show that $D^{(-)}_m(x)$ has no
canonical restriction to $\Sigma=\{x^0+x^3=0\}$. By Remark~\ref{rem:alt_sigma_res}
we have to show that $N_\lambda\cap\WF(D^{(-)}_m)\neq \emptyset$ where 
$N_\lambda$ is the set of normals of $\lambda:\R^3\to\R^4$, 
$(x^1,x^2,x^-)\mapsto(x^-/ \sqrt 2,x^1,x^2,-x^-/ \sqrt 2)$. One easily verifies that
$$
	N_\lambda=\{(x,\xi)\in\Sigma\times\R^4:\xi^0=\xi^3\}
$$
and hence $N_\lambda\cap\WF(D^{(-)}_m)\neq\emptyset$, which is easily seen by
considering Figure~\ref{fig:WF_neg_Pauli} and Figure~\ref{fig:NormalSetLambda}.
\end{proof}
So far we have shown that $D^{(+)}_m(x)$ and $D^{(-)}_m(x)$ have no canonical
restriction to $\Sigma=\{x^0+x^3=0\}$ -- notice that 
$\WF(D^{(+)}_m)=-\WF(D^{(-)}_m)$. Supplementary we will show that this also holds
true for the Pauli-Jordan function $D_m$ which is the sum of $D^{(+)}_m$ and
$D^{(-)}_m$.
\begin{proposition}\label{prop:Pauli-Jordan_no_restriction}
The Pauli-Jordan function $D_m(x)$ has no canonical restriction to $\Sigma$.
\end{proposition}
\begin{proof}
We will show that $\WF(D_m)=\WF(D^{(+)}_m)\cup\WF(D^{(-)}_m)$; the assertion
follows then from the proof of Theorem~\ref{thm:nonexistence_res_two-point}. Since
$D_m=D^{(+)}_m + D^{(-)}_m$ we get only one direction, namely
$\WF(D_m)\subset \WF(D^{(+)}_m)\cup\WF(D^{(-)}_m)$. 
To prove the other inclusion we may assume w.l.o.g.\ that 
$\WF(D_m)\cap\WF(D^{(-)}_m)\neq\emptyset$. Since $\cL^\uparrow_+$, the
restricted Lorentz group, operates transitively on $\WF(D^{(-)}_m)$ and since
$D_m$, and hence $\WF(D_m)$, is invariant under $\cL^\uparrow_+$, one obtains
$\WF(D^{(-)}_m)\subset\WF(D_m)$ (see also the proof of Thm. IX.48 in \cite{ReedII}).
Furthermore, $\WF(D^{(+)}_m)\subset\WF(D_m)$ since 
$\WF(D^{(+)}_m(x))=-\WF(D^{(-)}_m(x))\subset -\WF(D_m)$ and
$-\WF(D_m(x))=\WF(D_m(-x))=\WF(D_m(x))$.
\end{proof}
%
%------------------------------------------------------------------------------------------
\section{The tame restriction of the two-point function}
%------------------------------------------------------------------------------------------
%
The nonexistence of the restriction of the Pauli-Jordan function to $\Sigma=\{x^0+x^3\}$
is related to a fundamental problem in light cone quantum field theory where one describes
the dynamics of a quantum field by using $x^+=(1/\sqrt 2)(x^0+x^3)$ as ``time''-evolution
parameter. In this context it is essential to have well-defined fields for fixed $x^+=const.$.
However, to carry out the standard construction of a free field for fixed time, one has
to remain in a proper subspace of $\cS(\R^3)$ \cite{LKS} which was considered as a fault of
the theory \cite{SS}. In \cite{PullJMP} this problem was solved by introducing a new
test function space $\cS_{\partial_-}(\R^3)$ on which the ``restriction'' of the free field
can be defined and which determines the covariant field uniquely -- we called this
the ``tame restriction'' of the free field to $\Sigma$. Now, since the covariant commutator
relation of a free field $\phi$ reads
\begin{equation}\label{equ:covariant_commutator}
	[\phi(x),\phi(y)]=-iD_m(x-y),
\end{equation}
where $D_m$ is the Pauli-Jordan function, we see that the problem of nonexistence of
the real scalar field on $\Sigma$ results in the nonexistence of the restriction of
the Pauli-Jordan function to $\Sigma$. In \cite{Ull04sub} we introduced the tame restriction
of a generalized function and computed it for the Pauli-Jordan function, where we obtained
$(1/4)\delta(\bx_\bot)\otimes\epsilon(x^-)$. Hence, if we take the tame restrictions
(to $\Sigma$) on both sides of \eqref{equ:covariant_commutator} we arrive at the
well-known commutator relation of light cone quantum field theory \cite{LC-commutator}.
The same happens with the two-point function $W_2(x,y)=\bravac\phi(x)\phi(y)\vac$ 
since
\begin{equation}\label{equ:two-point}
	\bravac\phi(x)\phi(y)\vac = -i D^{(-)}_m(x-y),
\end{equation}
and $D^{(-)}_m$ does not have a canonical restriction to $\Sigma$.
However, since $D^{(-)}_m$
is a solution of the Klein-Gordon equation $(\Box+m^2)D^{(-)}_m=0$ we know from
\cite{Ull04sub} that $D^{(-)}_m$ admits a tame restriction to $\Sigma$.
In the sequel we will compute
this tame restriction explictly and show that it is independent of mass. 
Since the tame restriction of the free field to $\Sigma$ is also independent of mass 
\cite{PullJMP, PhD} no inconsistency appears if we take the tame restrictions (to 
$\Sigma$) on both sides of \eqref{equ:two-point}. First of all we have to recall the
definition of the tame restriction of a generalized function to $\Sigma$ -- for details see
\cite{PullJMP, Ull04sub}.
\begin{definition}
(a) Let $\cS_{p^+}(\R^n)=\bigcap_{k\geq 0}\{(p^+)^kg:g\in\cS(\R^n)\}$ be the topological
vector space endowed with the subspace topology induced by $\cS(\R^n)$; the dual space
$\cS'_{p^+}(\R^n)$ is called the space of 
{\em squeezed generalized functions}.

(b) Let $\cS_{\partial_-}(\R^n)=\bigcap_{k\geq 0}\{\partial^k_{x^-}g:g\in\cS(\R^n)\}$ be
the topological vector space endowed with the subspace topology induced by $\cS(\R^n)$;
the dual space $\cS'_{\partial_-}(\R^n)$ is called the space of 
{\em tame generalized functions}.
\end{definition}
The spaces $\cS_{p^+}(\R^n)$ and $\cS_{\partial_-}(\R^n)$ are Fr\'echet spaces.
Furthermore, the Fourier transform, which is an isomorphism from $\cS(\R^n)$ onto
$\cS(\R^n)$, maps $\cS_{\partial_-}(\R^n)$ onto $\cS_{p^+}(\R^n)$. Since we
are using light cone coordinates to represent $\Sigma$ (as $\{x^+=0\}$)
we also have to use the so-called $\LC$-Fourier transformation
$\cF_\LC:\cS(\R^n)\to\cS(\R^n)$ defined by
$$
	\cF_\LC(f)(\LCp)=\int f(\LCx)e^{i\LCbil{\LCx}{\LCp}}d\LCx,
$$
where $\LCbil{\LCx}{\LCp}=x^+p^- + x^-p^+ - \bx_\bot\cdot\bp_\bot$. Since $x^+$ is
the time variable in light cone physics we also introduce the spatial part of the
$\LC$-Fourier transformation
$$
	\pLCFT(f)(\LCbp)=\int f(\LCbx) e^{i(x^-p^+ - \bx_\bot\cdot\bp_\bot)}d\LCbx,
$$
and, in the special case of only one dimension,
$$
	\cF_\LC^{x^-\to p^+}(f)(p^+)=\int f(x^-) e^{ix^- p^+} dx^-.
$$
Clearly, $\cF_\LC$, $\pLCFT$ and $\cF_\LC^{x^-\to p^+}$ are isomorphisms from
$\cS(\R^x)$ onto $\cS(\R^x)$ which map $\cS_{\partial_-}(\R^x)$ onto
$\cS_{p^+}(\R^x)$ ($x$ appropriately chosen) which extend canonically to sequentially
continuous maps from $\cS'(\R^x)$ onto $\cS'(\R^x)$ respectively from
$\cS'_{\partial_-}(\R^x)$ onto $\cS'_{p^+}(\R^x)$.
\begin{definition}[Tame Restriction]
(a) A generalized function $\phi(y,z,x^-)\in\cS'(\R^{m+n+1})$ admits a tame restriction
to $\{y=y_0\}$ ($y_0\in\R^m$) 
if there is an open neighborhood $\Omega\subset\R^m$ of $y_0$ and a family 
$(\phi_y)_{y\in\Omega}$ with 
$\phi_y\in\cS'_{\partial_-}(\R^{n+1})$ ($y\in\Omega$) such that
$\Omega\to\C$, $y\mapsto(\phi_y,g)$ is $C^\infty$ for all $g\in\cS_{\partial_-}(\R^{n+1})$
and
$$
	(\phi(y,z,x^-),f(y)g(z,x^-))=\int_\Omega dy (\phi_y,g)f(y)
$$
for all $f(y)\in\cD(\Omega)$ and $g(z,x^-)\in\cS_{\partial_-}(\R^{n+1})$. In this case
we call $\phi\tres_{y=y_0}=\phi_{y_0}\in\cS'_{\partial_-}(\R^{n+1})$ the 
{\em tame restriction} of $\phi$ to $\{y=y_0\}$.

(b) A generalized function $\phi(x_1,\ldots,x_r)\in\cS'(\R^{4r})$ admits a tame restriction to
$\Sigma_{\tau_1}\times\cdots\times\Sigma_{\tau_r}$ 
($\Sigma_{\tau_i}=\{x_i\in\R^4:(1/\sqrt 2)(x_i^0 + x_i^3)=\tau_i\}$, $i=1,\ldots,r$) if
$\phi(\kappa^{-1}(\LCx_1),\ldots,\kappa^{-1}(\LCx_r))$ admits a tame restriction to
$\{x_1^+=\tau_1,\ldots,x_r^+=\tau_r\}$; in this case we call 
$\phi\tres_{\Sigma_{\tau_1}\times\cdots\times\Sigma_{\tau_r}}
=\phi(\kappa^{-1}(\LCx_1),\ldots,\kappa^{-1}(\LCx_r))%
\tres_{x_1^+=\tau_1,\ldots,x_r^+=\tau_r}$
the {\em tame restriction} of $\phi(x_1,\ldots,x_r)$ to 
$\Sigma_{\tau_1}\times\cdots\times\Sigma_{\tau_r}$.
\end{definition}
\begin{proposition}\label{prop:neg_freq_Pauli-Jordan_tame_res}
Let $D^{(-)}_m(x)\in\cS'(\R^4)$ denote the negative-frequency Pauli-Jordan function.
Then $D^{(-)}_m(x)$ admits a tame restriction to $\Sigma_\tau$ 
($\tau\in\R$) and
%$$
%	D^{(-)}_m\tres_\Sigma=i \delta(\bx_\bot)\otimes G(x^-),
%$$
%where $G=(\cF_\LC^{x^-\to p^+})^{-1}(\Theta(p^+)/p^+)\in\LCcS'(\R)$,
%i.e., 
$$
	(D^{(-)}_m\tres_{\Sigma_\tau},g)=\frac{-1}{i(2\pi)^3}\int_{p^+<0}
		\frac{d^3\LCbp}{2|p^+|}(\pLCFT{g})(\LCbp)e^{i\LComega(\LCbp)\tau}.
$$
for all $g(\LCbx)\in\cS_{\partial_-}(\R^3)$
\end{proposition}
\begin{proof}
Let $f(x^+)\in\cS(\R)$ and $g(\LCbx)\in\cS_{\partial_-}(\R^3)$. By definition
\begin{align}
	((D^{(-)}_m\circ\kappa^{-1})(x^+,\LCbx),f(x^+)g(\LCbx)) & =
	\frac{-1}{i(2\pi)^3}(\delta_-(\LCp^2-m^2),\hat f(-p^-)
	\LChat g(\LCbp))\nonumber \\
	& = \frac{-1}{i(2\pi)^3}\int_{p^+<0}\frac{d^3\LCbp}{2|p^+|}
	\hat f(-\LComega(\LCbp))\LChat g(\LCbp)\label{eq:in0}
\end{align}
Since $g\in\cS_{\partial_-}(\R^3)$ we have
$f(x^+)\LChat g(\LCbp)\in
\cL^1(\R\times\R^3,dx^+\otimes\frac{d^n\LCbp}{|p^+|})$. Hence we can put
$\hat f(-\LComega(\LCbp))=\int dx^+ f(x^+)e^{i\LComega(\LCbp)x^+}$ in
\eqref{eq:in0}, and obtain
$$
	((D^{(-)}_m)_{x^+},g)=\frac{-1}{i(2\pi)^n}\int_{p^+<0}
	\frac{d^n\LCbp}{2|p^+|}\LChat g(\LCbp)e^{i\LComega(\LCbp)x^+}.
$$
Thus the assertion follows since $(D^{(-)}_m\tres_{\Sigma_\tau},g)=((D^{(-)}_m)_\tau,g)$.
\end{proof}
\begin{remark}
Notice that $D_m^{(-)}$ is uniquely determined by its tame restriction
to $\Sigma_0$ \cite{Ull04sub}.
\end{remark}
\begin{remark}\label{rem:tres_diff}
One can easily verify that if a generalized function $\psi(x,y)\in\cS'(\R^4\times\R^4)$ is of
the form $\psi(x,y)=\phi(x-y)$, where $\phi\in\cS'(\R^4)$, and $\phi$ has a tame restriction
to $\tau_1 - \tau_2$ then $\psi$ has a tame restriction to
$\Sigma_{\tau_1}\times\Sigma_{\tau_2}$ and
$\psi\tres_{\Sigma_{\tau_1}\times\Sigma_{\tau_r}}=
\phi\tres_{\Sigma_{\tau_1 - \tau_2}}(\LCbx - \LCby)$. Notice that
$(\phi(x-y),f(x)g(y))=(\phi,f* g^\lor)$, where ``$*$'' means convolution and
$g^\lor(x)=g(-x)$.
\end{remark}

\begin{corollary}
Let $\phi(x)$ be the real scalar free field of mass $m>0$, and 
$W_2(x,y)=\bravac\phi(x)\phi(y)\vac$ the associated two-point function. 
Then $W_2(x,y)$ admits a tame restriction to 
$\Sigma_\tau\times\Sigma_\tau=\{x^+=y^+=\tau\}$ ($\tau\in\R$) and
$$
	W_2(x,y)\tres_{\Sigma_\tau\times\Sigma_\tau}=
	\delta(\bx_\bot - \by_\bot)\otimes G(x^- - y^-)\in\cS'_{\partial_-}(\R^3\times\R^3),
$$
where $G=(\cF_\LC^{x^-\to p^+})^{-1}(\Theta(p^+)/p^+)\in\LCcS'(\R)$.
In particular, the tame restriction of $W_2(x,y)$ to $\Sigma_\tau\times\Sigma_\tau$ is
independent of mass.
\end{corollary}
\begin{proof}
Since $W_2(x,y)=-iD^{(-)}_m(x-y)$ it is enough to show that $D^{(-)}_m$ admits
a tame restriction to $\Sigma=\{x^+=0\}$ and that 
$D^{(-)}_m\tres_\Sigma=i\delta(\bx_\bot)\otimes G(x^-)$
(cf.\ Remark~\ref{rem:tres_diff}); however, this follows immediately
from Proposition~\ref{prop:neg_freq_Pauli-Jordan_tame_res}.
\end{proof}
%
%-----------------------------------------------------------------------------
\section{Conclusion}
%-----------------------------------------------------------------------------
%
To get rid of the (perturbative) zero-mode and restriction problem in light cone quantum 
field theory, we have introduced in \cite{PullJMP}
the function space $\LCcS(\R^3)$ and its dual space -- the space of tame
generalized functions. The restriction problem, i.e., the
problem that the real scalar free massive field has no canonical restriction to
$\Sigma=\{x^0+x^3=0\}$, manifests itself in the problem that the 
(positive-/ negative-frequency) Pauli-Jordan  has no canonical restriction to $\Sigma$ 
in the sense of distribution theory.
By using the so-called tame restriction of a tempered distribution, which we have already
introduced in \cite{Ull04sub}, we have seen that also
the assumed inconsistency of the mass-dependence of the two-point function on
$\Sigma$ can be resolved. Thus the result of this paper contributes to the 
philosophy (introduced in \cite{PullJMP}) that $\cS_{\partial_-}(\R^3)$ 
-- instead of $\cS(\R^3)$ -- is the right test function space
when treating quantum fields on the null plane $\Sigma$. 
%-----------------------------------------------------------------------------


\begin{thebibliography}{99}

%\bibitem{ComAlg} Atiyah, M.F., MacDonald, I.G.: {\em Introduction to
%Commutative Algebra}. Addison-Wesley, (1969)

\bibitem{Bog1} Bogolubov, N.N. et al.: {\em General principles of
quantum field theory}. Kluwer Academic Publisher (1990).

\bibitem{BrosPins} Brodsky, S.J., Pauli, H.-C.:
Quantum Chromodynamics and Other Field Theories on the
Light Cone. Phys. Lett. C \textbf{301}, 299 (1998), hep-ph/9705477.

\bibitem{LC-commutator} Chang, S., Root, R.G. and Yan, T.: Quantum field theories in the
infinite-momentum frame. I. Quantization of scalar and Dirac fields.
Phys. Rev. D  \textbf{7},  1133 (1973).

%\bibitem{Dirac} Dirac, P.A.M.: Forms of relativistic dynamics. Rev. Mod. Phys. \textbf{21},
%392 (1949)

\bibitem{Ehrenpreis1962} Ehrenpreis, L.: Solution of some problems of division,
Part IV. American Jour. of Math. \textbf{82}, 522 (1962).

\bibitem{Garding1958}
G\aa rding, L., Malgrange, B.: Op\'erateurs diff\'erentiels
partiellement hypoelliptiques. C. R. Acad. Sci. \textbf{247},
2083 (1958).

\bibitem{Garding1961}
G\aa rding, L., Malgrange, B.: Op\'erateurs diff\'erentiels
partiellement hypoelliptiques et partiellement elliptiques. Math. Scand.
\textbf{9}, 5 (1961).

%\bibitem{Hartshorne} Hartshorne, R.: {\em Algebraic Geometry}, Springer-Verlag,
%Berlin (1977)

%\bibitem{HeinzlWerner} Heinzl, Th., Werner, E.: Light-front quantization as
%an initial-boundary-value problem. Z.\ Phys.\ C \textbf{62}, 521 (1994)

\bibitem{Hoer1} H\"ormander, L.: {\em The analysis of linear partial
            differential operators I}. Springer-Verlag,  Berlin (1990).

%\bibitem{Hoer2} H\"ormander, L.: {\em The analysis of linear partial
%            differential operators II}. Springer-Verlag,  Berlin (1990)

\bibitem{LKS} Leutwyler, H., Klauder, J.R. and Streit L.:
Quantum field theory on lightlike slabs. Nuovo Cimento A  \textbf{66}, 536 (1970).

%\bibitem{LeutStern} Leutwyler, H. and Stern J.: Relativistic Dynamics on
%a Null Plane. Ann. Phys. \textbf{112}, 94 -- 164 (1978)

\bibitem{NakYam77} Nakanishi, N., Yamawaki, K.: A consistent formulation of the null-plane
quantum field theory, Nucl. Phys. \textbf{B122}, 15 (1977).

%\bibitem{ReedI} Reed, M. and Simon, B.:
%    {\em Methods of modern mathematical physics I}. Academic Press (1975)

\bibitem{ReedII} Reed, M. and Simon, B.:
 {\em Methods of modern mathematical physics II}. 
Academic Press, New York (1975).

\bibitem{Rudin_func} Rudin, W.: {\em Functional Analysis}. McGraw Hill,
Reprint (1990).

\bibitem{SS} Schlieder, S. and Seiler E.: Some Remarks on the Null
Plane Development of a Relativistic Quantum Field Theory.
Commun. Math. Phys. \textbf{25}, 62 (1972).

%\bibitem{Schweber} Schweber, S.S.: 
%{\em An Introduction to Relativistic Quantum Field Theory}.
%Harper \& Row (1962)

\bibitem{PullJMP} Ullrich, P.:
On the restriction of quantum fields to a lightlike surface. J. Math. Phys. \textbf{45},
3109 (2004).

\bibitem{Ull04sub} Ullrich, P.: Uniqueness in the characteristic Cauchy problem of the
Klein-Gordon equation and tame restrictions of generalized functions. Submitted (2004),
math-ph/0408022.

\bibitem{PhD} Ullrich, P.: A Wightman approach to light cone quantum field theory.
Preprint (2004).

%\bibitem{Pull_future1} Ullrich, P.:
%Tame Restrictions of Distributions and the Wave Front Set. In preparation

%\bibitem{Pull_future2} Ullrich, P.: $F$-tempered distributions. In preparation.

\bibitem{Yam97} Yamawaki, K.: Zero Mode and Symmetry Breaking on the Light Front,
Proc. of Int. Workshop {\em New Nonperturbative Methods and 
Quantization on the Light Cone}, Les Houches, France (1997),
hep-th/9707141.

\end{thebibliography}
\end{document}